# Intraseasonal Oscillation of Land Surface Moisture and its role in the maintenance of land ITCZ during the active phases of the Indian Summer Monsoon

Pratibha Gautam1,2, Rajib Chattopadhyay1,3*, Gill Martin4, Susmitha Joseph1, A.K. Sahai1,

1. Indian Institute of Tropical Meteorology, Pune

2. Savitribai Phule Pune University

3. India Meteorological Department Pune, Maharashtra 411005

4. Met Office, Exeter, UK

*Corresponding Author

Rajib Chattopadhyay

Indian Institute of Tropical Meteorology,

Dr. Homi Bhabha Road, Pune 411008, India

rajib@tropmet.res.in

# Abstract

What is the role of soil moisture in maintaining the land ITCZ during the active phase of the monsoon? This question has been addressed in this study by using ERA5 reanalysis datasets, and then we evaluate the question in the CFS model-free run. Like rainfall, soil moisture also show intraseasonal oscillation. Furthermore, the sub-seasonal and seasonal features of soil moisture are different from each other. During the summer monsoon season, the maximum soil moisture is found over western coastal regions, central parts of India, and the northeastern Indian subcontinent. However, during active phases of the monsoon, the maximum positive soil moisture anomaly was found in North West parts of India. soil moisture also play a pre-conditioning role during active phases of the monsoon over the monsoon core zone of India. When it is further divided into two boxes, the north monsoon core zone, and the south monsoon core zone, it is found that the preconditioning depends on that region's soil type and climate classification. Also, we calculate the moist static energy (MSE) budget during the monsoon phases to show how soil moisture feedback affects the boundary layer MSE and rainfall. A similar analysis is applied to the model run, but it cannot show the realistic preconditioning role of soil moisture and its feedback on the rainfall as in observations. We conclude that to get proper feedback between soil moisture and precipitation during the active phase of the monsoon in the model, the pre-conditioning of soil moisture should be realistic.

# 1. Introduction

The Indian summer monsoon season experiences a double Inter-Tropical Convergence Zone (ITCZ): one over the ocean and one over the land (Gadgil & Srinivasan, 1990). However, these zones of convergence, cloudiness, and rainfall are not simultaneously visible on a daily chart. Instead, they vary in a well-known see-saw pattern. The land ITCZ shows intraseasonal oscillations (ISO) and is visible when the northward propagating ISO phase is over land (Sikka and Gadgil, 1980). The rainfall over the core monsoon zone is more prominent than climatology during this period (i.e., there is an active phase). Oceanic ITCZ is driven by classical SST-moisture-convection feedback (Neelin & Held, 1987). Despite the ITCZ being the same magnitude of strength over the land region as over the ocean, it is not well known how or whether land surface moisture-convection feedback is essential for the growth and maintenance of the land ITCZ. Neither is it well known what is the nature of the land surface-to-land ITCZ feedback over the core monsoon zone during the active phase. Intense diabatic heating maintains the land ITCZ (Chattopadhyay et al., 2013) once the ITCZ is established over land. However, diabatic heating depends on convective scale systems with a low lifetime (~hrs), while the typical active phase lasts for a few days. Intuitively, intraseasonal maintenance of the land ITCZ requires long memory feedback from slowly varying processes. What is the source of such low-memory feedback? Over the ocean, intraseasonal waves such as MJO or the summer intraseasonal mode get ample moisture feedback, now known as moisture pre-conditioning (Chen & Zhang, 2019; Sobel & Maloney, 2013). Theories predict that moisture pre-conditioning is extremely important for maintaining MJO or ISO. Sea surface temperature and evaporative fluxes over the ocean and atmospheric boundary layer are essential regulators of moisture convection feedback.(Bladé & Hartmann, 1993; Li et al., 2018; Sabeerali et al., 2014; Sharmila et al., 2013)

Are there any such local moisture-driven mechanisms for the land ITCZ and active phases of the monsoon? Several studies have shown that the land surface processes control the timing and intensity of convection over different parts of the world (De Ridder, 1997; Ganeshi et al., 2020; Niyogi et al., 2010; Pielke, 2001; Taylor et al., 2012). Several mechanisms are suggested for how the land surface controls the moisture-convection feedback over land. One of the famous mechanisms was proposed by Eltahir, 1998 and Seneviratne et al., 2010. In that hypothesis (Eltahir, 1998), if we have wet soil moisture there are mainly two ways to affect the rainfall: one by changing the Bowen ratio and another through the surface albedo. And these affect the moist static energy in the atmospheric boundary layer and the rainfall pattern over that region. Again, the feedback starts from rainfall to soil moisture. Over the Indian region, several studies suggest the role of soil moisture memory in the seasonal monsoon and intraseasonal forecast (Halder et al., 2018; Koster et al., 2002, 2006, 2011; MOHANTY et al., 2021; Saha et al., 2012; Unnikrishnan et al., 2017) . Menon et al., (2018) shows how shallower convection helps to precondition the atmosphere for deeper convection during the onset progression, and Menon et al. (2022) further illustrate that the specified soil and vegetation conditions affect the timescales of soil moisture–precipitation feedback and thereby the local onset through the timing of diurnal convection.

The studies mentioned above are not well understood if the land surface parameters, like soil moisture at different levels, and surface temperature, play an essential role in the intraseasonal scale organization of land ITCZ during the active phase. As a first glimpse of why it could be significant, we show Fig.1 . We have shown the JJAS seasonal climatology of soil moisture and mean sea-level pressure from ERA5 data, and the active phase composite of anomalous soil moisture, mean sea level pressure, and the vertical velocity at 500 hPa. Although for rainfall, the plot shows a similar spatial distribution, which is expected as seasonal and intraseasonal modes show a standard common mode of variability (Goswami &

Mohan, 2001), the anomalous soil moisture distribution shows an opposite gradient in the active phase composite Fig.1(b) as compared to the seasonal mean composite Fig.1(b). The seasonal soil moisture pattern decreases from central India towards northwest India, while during the active phase, the soil moisture anomaly increases towards northwest India from central India. This reversal in a soil–moisture anomaly gradient within the core monsoon zone in the active composite compared to the seasonal mean composite is intriguing as it suggests a reduction in the west to east soil moisture gradient. This is also confirmed using other model data based on the Monsoon Mission project (Nayak et al., 2018), and the results are consistent in terms of the change in gradient (although the actual spatial pattern of soil moisture amplitude differs, which may be due to a difference in soil depth). The change in this gradient provides a first-hand glimpse that the intraseasonal feedback during the active phase could be one crucial driving factor in the preconditioning process in the western part of the monsoon zone and is different compared to the eastern part of the monsoon zone. Thus, a difference in feedback could occur within the core monsoon zone.

The gradient change, however, could be related to soil type and land surface vegetation. The eastern side and western side are made of different types of soil (Fig.1c). Increased soil moisture over the western alluvial and black soil as compared to the eastern red soil is expected as it is well known that alluvial and black soil has high water retention capacity. In contrast, like sandy soil, red soil has a low water retention capacity (Lehmann et al., 2018), leading to less soil moisture availability. Such soil type difference provides differences in land surface water budget and storage (or matric potential) terms. Also, the monsoon zone falls under two different Kopens climate classification zones (Raju et al., 2013). Thus, propagating northward, intraseasonal oscillations would face different land surface feedback over the monsoon zone, and the structure of the ITCZ feedback will be different. Pangaluru et al., (2019), using a maximum covariance analysis, showed a difference in monthly soil moisture and total water storage spatial location (fig.7 of their paper). Similarly, they have shown that, on the monthly scale, rainfall and soil moisture have a spatial gradient in distribution.

Here we have shown that spatial anomaly gradients occur on intraseasonal timescales in the active composites. Menon et al. (2022) showed that, during monsoon onset, changing the specification of the soil type in a convection-permitting model affected the spatial variations in latent and sensible heat fluxes, altering the diurnal cycle of convection and the convective initiation along the leading edge of the monsoon progression. A relevant question is whether the land surface takes part in the moisture feedback preconditioning on the intraseasoonal timescale, as is well-known in oceanic surface preconditioning for the oceanic ITCZ. The identification of the land-surface-to-land-ITCZ feedback during the active phase is thus necessary. The necessity of the current analysis also arises from the modeling perspective. It is also well known that the operational forecast models show dry bias over land (Abhilash et al., 2014). So far, it is not clear from studies if the nature of feedback is represented appropriately in the forecasting model and if incorrect feedback contributes to the dry bias. Several studies (Baisya et al., 2017; Hunt et al., 2016; Osuri et al., 2020; Rajesh et al., 2017) show the track of monsoon depression is impacted by additional soil moisture over the propagation tracks of the depressions. The land surface significantly influences extremes, such as drought and heat waves (Roundy et al., 2013; Roxy et al., 2017). Therefore, understanding the nature of low-frequency intraseasonal feedback is also essential for improvements in monsoon modelling (Webster, 1983).

Based on the above review, thus, the objectives of the current study are: (a) To bring out the intraseasonal fluctuation of soil moisture and show the expected nature of feedback during the active phase over the monsoon zone, (b) To evaluate the observed feedback in the operational models. The evaluation will be based on free (uninitialized) model runs and operational forecasts of active spells.

## 2. Data and Methodology

The present research uses ERA5 reanalysis datasets (Hersbach et al., 2020) (https://www.ecmwf.int/en/forecasts/datasets/reanalysis) for soil moisture at level 1 (0-10cm), level 2 (10-28cm), level 3 (29-78), and level 4 (79-100cm), the same for soil – temperature data at different levels along with evaporation, specific humidity and temperature data for the period 1989-2019. NCEP and Nayak et al. (2018) datasets are also used for soil–moisture parameters, but the soil–moisture level is different in all the data sets. In Nayak et al. (2018), soil moisture level 1 is 0-7cm. The Nayak et al. (2018) data is available from the Monsoon Mission project (https://www.tropmet.res.in/monsoon/monsoon2/). Hereafter we denote this data as MMSM (Monsoon Mission Soil Moisture). In addition to the reanalysis soil moisture data, we used IMD (India Meteorological Department) gridded rainfall data from Pai et al. (2014). The gridded rainfall data are available for the entire Indian subcontinent (6.5N–37.5N and 66.5E–101.5E).

We identify active and break days in this study using Pai et al. (2014) gridded rainfall data. Active days are defined when the standardized rainfall anomalies over India's monsoon core zone exceed 1 based on rainfall data filtered to 20-100 days. Similar criteria are considered for model-free runs.

Day 0 represents the maximum rainfall averaged over the monsoon core zone. Also, a power spectrum analysis has been performed to show the intraseasonal variability of rainfall, evaporation, and soil moisture and calculate the standardized anomaly:

Standardized anomaly = (anomaly from climatology)/standard deviation

## 3. Results

### 3.1. Seasonal and sub-seasonal features of Soil moisture

It is already known that for rainfall, the seasonal and intraseasonal modes show a standard common mode of variability (Goswami & Mohan, 2001). But in the case of soil moisture, we found it is different. We have investigated the seasonal and sub-seasonal soil moisture features at soil level 1.

The seasonal and sub-seasonal variations of soil moisture show different characteristics. Fig 1 (a) shows the JJAS (June to September) climatology of soil moisture level 1 (level 1, 0-10cm, shaded) and mean sea level pressure (hPa, contour). Fig 1(b) shows the active days composite of anomalous soil moisture lev1 obtained from ERA5 (shaded), mean sea level pressure (contour), and dotted represent where there are substantial positive anomalous values for omega at 500hPa. During the summer monsoon season, the maximum soil moisture is found over western coastal regions, central parts of India, and the northeastern Indian subcontinent. However, during the active phase of the monsoon (i.e., a sub-seasonal feature of monsoon), the maximum positive soil moisture anomaly was found in North West parts of India. In Fig 1(a), the climatological pattern of mean sea level pressure (MSLP) shows an extended area of low pressure in northwestern India, which covers the entire northern part of India, the Indo-Gangetic plains, and south of the Himalayas with values below 1002 hPa. During the active phase of the monsoon in fig 1(b), MSLP shows a similar JJAS climatological pattern with lower values below 1000hPa, and the dotted pattern shows the positive anomalous values of omega at 500 hPa during the active phase of Monsoon over India's south monsoon core zone. Some positive values also show over the North of India. Fig 1(c) and 1(d) show the different soil types of India and the Koppen climate classification of the Indian region from 1981-2010. Different soil types or climate classifications have different responses to rainfall in the soil parameters. For example, an arid climate with sandy soil will have a different response of soil moisture to rainfall than an alluvial or black soil. Sandy soil has more spaces and less easily bind to water molecules, hence more free water is available. This can be explained in terms of soil matric potential which measures the availability of free water. Thus, the spatial pattern of rainfall and soil moisture is likely to differ for the same amount of rainfall..

Next, we examine the temporal variations. As we know, rainfall has a low-frequency variability signal. Now to see whether the soil parameter also shows the intraseasonal variation, we plot fig 2, which shows the intraseasonal oscillation for various surface parameters from ERA5 datasets. Fig 2(a), (b), and (c) shows the time series of rainfall, evaporation, and soil moisture at lev1 for JJAS of the year 2000. Fig 2(d),(e), and (f) shows their power spectra respectively over Central India for the period (1989-2019). So from the time series and power spectra plot, we can say that the rainfall, evaporation, and soil moisture all show the maximum peaks in the low-frequency band (30-45) days on 95% significance levels. MMSM datasets also show the low-frequency variability in soil–moisture datasets, shown in supplementary fig S1(a, b). Thus, the analysis shows strong sub-seasonal variability of soil moisture and rainfall. We further performed a correlation analysis during the JJAS summer monsoon between both the unfiltered (fig 3) and filtered (supplementary fig S2) rainfall and soil moisture parameters at different levels (levels -1, 2,3 and 4) over Central India (1989-2019). It is found that there is a positive relationship between rainfall and soil moisture in the uppermost soil levels. Maximum correlation occurs at soil level -1(r=0.63); after that, it reduces in levels 2, 3, and 4. Thus, this section shows that (a) the spatial pattern variation of rainfall and soil moisture are not uniform over the Indian region and (b) there is a strong intraseasonal variation of rainfall and soil moisture with level-1 soil moisture showing an in-phase relationship.

## 3.2 Spatially asymmetric propagation of moisture and identification of pre-conditioning role

Previous studies have shown that moisture pre-conditioning is extremely important for maintaining ISO and sea surface temperature, and evaporative fluxes over the ocean and atmospheric boundary layer are essential regulators of moisture convection feedback. But over the core monsoon zone, are there any such mechanisms for land ITCZ and active phases of monsoon? This is explored next. Fig 4 depicts the latitude sections, averaged over 68°-96°E, of daily soil moisture anomalies at different levels, lagged with respect to rainfall maxima (at lag 0), for the period 1989-2017. If we consider soil level-1(SM-1) in fig 4(a), we can see that the there is northward propagation of soil moisture and rainfall anomalies. At deeper soil levels the soil moisture lags the rainfall by increasing numbers of days.

as seen in fig 5 where anomalous soil moisture and rainfall propagation are shown during the active phase of monsoon with lead-lag 12 days around the peak active day (day 0). Apparent northward propagation of soil moisture (shaded) is seen in fig 5 (we also show a similar plot from MMSM data in supplementary fig S3) as we go towards day zero (i.e., the peak active phase of the monsoon). It starts from the south tip of the Indian subcontinent and then moves towards the north. It is important to note that although the start of the propagation of soil moisture and rainfall from the southern part of India is the same, the soil moisture and rainfall propagation have changed when it moves toward India's monsoon core zone. We can see that the soil moisture anomaly reaches the north of the monsoon core zone earlier than the rainfall. For this clarification, we have divided the monsoon core zone into two parts: the north box of the monsoon core zone and the south of the monsoon core zone, as shown in Fig 6.

Fig 6 (a) represents the spatial pattern of anomalous SM-1 and rainfall during the peak active composite days (day 0). To check the propagation of soil moisture over the monsoon core zone in India, we split the monsoon core zone into two boxes north of the monsoon core zone (N) and south of the monsoon core zone (S), which is shown in fig 6(a). To investigate the moisture pre-conditioning over these two boxes, we have taken the difference between anomalous soil moisture and rainfall over these two boxes. Fig 6(b) shows the plot of the standardized SM-1 and rainfall anomaly gradient over the (Northern - Southern) monsoon core zone during active days from ERA5, NCEP, and MMSM datasets. Positive values on the y-axis show the increase in the north-south gradient of anomalous soil moisture and rainfall, i.e., the value of the variables in the northern box is more significant than the value of the variables in the southern box. However, it may be noticed that there is a temporal shift in the soil moisture and rainfall anomaly gradient over India's north and south boxes inside the monsoon core zone: soil moisture increases earlier over the northern box in the western part of India, with increased soil moisture occurring a few days before the increase in rainfall is visible. So from figs 4,5, and 6, we can say that soil moisture may have a pre-conditioning role during the monsoon's active phase. Also, fig 6(c) and fig (d) shows the probability distribution of standardized rainfall and soil moisture anomaly over the north and south boxes of the monsoon core zone for days -5 to -1 before the peak active day. Both boxes show a similar distribution for the rainfall, but significant change occurs in soil moisture distribution, with larger soil moisture anomalies in the north box.

It is also found that this moisture availability is linked with the near-surface variables (i.e., evaporation). Fig 7 shows the spatial distribution of actual evaporation (m/day, shaded) and mean sea level pressure contours in the days leading up to the peak active rainfall day. Since the ERA5 convention is that downward fluxes are positive, we have multiplied the values by minus 1 for our convenience to understand that positive values represent increased evaporation. It is interesting to note from fig 7 that as we go towards day 0, i.e., peak active days, evaporation is increasing in North West of India (from Rajasthan), and the contour of minimum low-pressure area is expanded from northwest India. So we can say that soil moisture affects near-surface variables that can influence precipitation, which is already known in previous studies.

Further, to know the impact of soil moisture on the vertical moisture profile, we have shown in fig 8 the vertical profile of the N minus S monsoon core zone difference in specific humidity anomaly during the active phase of the monsoon. Positive color index (1,2,3,4,5) represents lags from day 0, and (-1,-2,-3,-4,-5) represents lead days up to day 0. Fig 8 shows that the specific humidity anomaly gradient is positive in the boundary layer. The change in the anomalous specific humidity gradient is larger on the days before the rainfall maximum (indicated by a blueish color shade), while it is small on the days after the rainfall maximum.

Along with the vertical moisture profile, we have also investigated the soil-moisture effect on moist static energy (MSE), a measure of intraseasonal oscillation. Studies (Eltahir, 1998) have shown the impact of soil moisture on MSE through its feedback process. We have investigated soil moisture's role on MSE during the active monsoon phase. Fig 9 shows the spatial distribution of anomalous MSE (shaded) averaged over 1000-150hPa during the active phase (peaking at lag 0). Rainfall (mm/day) is shown as contours. From fig 9, we can see the northward propagation of anomalous MSE and that it increases in northwest India from around 8 days before the rainfall maximum, consistent with the changes in soil moisture.

To see how the atmospheric stability is related to MSE and further the monsoon rainfall, we have calculated the Brunt-Vaisala (BVF) frequency (1/s), which is a measure of instability, shown in fig 10 with 10 days lead-lag around the day of the peak active phase (i.e., day 0). The black line shows the BVF from ERA5 over the north box and the red color over the south box. The more negative values indicate more instability in the atmosphere. The ERA5 data in fig 10 show increasing instability over the north box in the days before the peak active phase. After day 0, the instability decreases gradually. And in the south box, instability starts to decrease three days before day 0. So we can say that instability develops first over India's north box of the monsoon core zone, before that over the south box.

Based on the above discussion, the suggested mechanism is summarized in fig 11. As per this mechanism, the spatially inhomogenous distribution of soil type over India's monsoon core zone will create a spatially asymmetric distribution of soil moisture anomalies: in sandy or arid soil type, for a small amount of rainfall a week before the peak active days, more water is freely available for evaporation or plant uptake (and hence transpiration) than in loam. Thus over the western side of the monsoon zone, moisture availability is increased more than on the eastern side a few days before the active phase, which causes the soil moisture anomaly asymmetry. Asymmetric soil moisture anomalies over the monsoon core zone will further affect the evaporation in an asymmetric way, which affects the organization of the monsoon trough, shown in fig 7. And spatial and temporal shifts in soil moisture will lead to more humidity in the atmosphere from the western side, which is clear from fig 8. This affects the MSE and organization of land ISO, which will affect the organized land ITCZ and rainfall.

### 3.3 Moisture Feedback from the CFS model free run

To see how an operational model performs during active days and whether the feedback mechanism exists, we have done the same analysis with the CFS free long run available in ERPAS (Extended Range Prediction) group. Fig 12 shows the anomalous soil moisture and rainfall propagation during the active phase of the monsoon with lead-lag 12 days from the CFS free run. From fig 12, we can say that no such gradients of soil-moisture anomalies exist in the model as were observed in the ERA5 reanalysis. Also, no such lagged relationship exists in fig 13 as was seen in fig 6b, which shows the evolution of standardized soil moisture and rainfall anomalies. No spatial and temporal shift occurred in the model free run. So there appears to be no feedback loop between soil moisture and rainfall in the model. One possible reason could be the improper distribution of soil -type and soil texture in the model (e.g., Lehmann et al., (2018) for the Gujarat region).

Further, soil moisture's role on the vertical moisture profile in the model-free run was investigated in fig 14. This shows that the model could not capture the observed evolution of specific humidity. We have seen in the observed moisture vertical profile difference pattern (fig 8) that the specific humidity anomaly gradient is positive in the boundary layer and that the change in the anomalous specific humidity gradient is more significant before the rainfall maximum (day 0) than after the rainfall maximum. However, in the case of the model-free run, the specific humidity anomaly gradient is negative in the boundary layer until 3 days before the rainfall maximum, and the difference in the rate of change before and after day 0 is smaller. Fig 10 shows that the model's Brunt Vaisala frequency (BVF) is entirely different from the observed pattern. Although the model also shows instability during the active phase because it has negative values, it is not the same as what we are getting in the observed pattern. This is likely because soil moisture and rainfall feedback during the active phase is improper in the model.

## 4. Conclusions

This study explored the intraseasonal variation of land surface moisture and its role in the maintenance of land ITCZ during the active and break phases of the Indian Summer Monsoon from ERA5 reanalysis datasets and verified the results with other available datasets. The results were then compared against the CFS model free run. Studies have shown that for rainfall, the seasonal and intraseasonal modes show a standard common mode of variability (Goswami & Mohan, 2001). But in the case of soil moisture, we found a weakening of the sub-seasonal gradient and a change in the spatial pattern compared to the seasonal gradient. The above analysis showed that soil moisture distribution's sub-seasonal and seasonal features differ. During the summer monsoon season, the maximum soil moisture is found over western coastal regions, central parts of India, and the northeastern Indian subcontinent fig 1(a). However, during the active phase of the monsoon (i.e., sub-seasonally), the maximum positive soil moisture anomaly was found in northwest parts of India fig1 (b). Like rainfall, land surface parameters (soil moisture and evaporation) also show intraseasonal oscillations, as shown in fig 2. There is a positive relationship between upper-level soil moisture and rainfall during the JJAS (1989-2019) season, which is seen in fig 3 and is mostly related to the intraseasonal timescale (fig. S3). Maximum correlation occurs between rainfall and soil moisture during JJAS over Central India in soil level -1(r=0.63); after that, it reduces in levels 2, 3, and 4.

Theories predict that moisture pre-conditioning is extremely important for maintaining ISO and sea surface temperature, and evaporative fluxes over the ocean and atmospheric boundary layer are essential regulators of moisture convection feedback. But over land, are there any such mechanisms for the land ITCZ during active phases of monsoon? This is explored here by analyzing the soil moisture features during active phases of monsoon. From the Hovmöller and spatial plots of soil moisture (figs 4 and 5), it is clear that land surface parameters (soil moisture) play a pre-conditioning role during the active phase of the monsoon over the monsoon core zone of India. Dividing the monsoon core region into two boxes suggests that preconditioning depends on that region's soil type and climate classification. Fig 6, in the lagged time series plot of soil moisture and rainfall anomaly gradient (north box minus south box) over the monsoon core zone, shows that the soil moisture anomaly gradient reaches its maximum before the rainfall anomaly gradient maximum. This suggests that soil moisture may act as a preconditioner during the active phase of the monsoon.

Soil moisture is linked with the near-surface variables (i.e., evaporation) through its feedback process. Fig 7 shows that as we go towards day 0, i.e., the peak active day, evaporation increases in the North West of India (from Rajasthan), and the contour of the minimum low-pressure area is also expanding from northwest India towards the Indo-Gangetic plain. Further, fig 8 shows that the specific humidity anomaly gradient between the north and south monsoon core zone is positive in the boundary layer and that the change in the anomalous specific humidity gradient is more considerable before than after the rainfall maximum. The changes in gradient between north and south of the monsoon core zone are largely driven by the increased soil moisture, evaporation and specific humidity in the north box during the active phase, since this region is usually relatively dry, whereas in the south box there is less increase (or even a decrease in evaporation) during the active phase because this region is usually moist.

Soil moisture also affects the moist static energy (MSE), a measure of intraseasonal oscillation. Studies by Prasanth and Sahai (2013) showed vertically integrated moisture and moist static energy (MSE) budgets during the active and break phases of the Indian summer monsoon. From fig 9, we can say that MSE increases around 8 days before the rainfall maxima from the northwest of India, which we also attribute to feedback from the soil moisture changes. We note, however, that there will also be atmospheric moisture advection ahead of the rain band that will contribute to the specific humidity changes. Menon et al. (2018, 2022) and Parker et al. (2016) showed how shallow convection helps to precondition the atmosphere for deeper convection during monsoon onset progression. It could be that, during intraseasonal oscillations, a similar process occurs (which is shown in S4 and S5 ) in which shallower convection produces light rainfall that, combined with the variation in soil types, causes the additional increase in soil moisture in the northwest, prompting the positive land-atmosphere feedback. This will be investigated in future work. In figure S6, we have shown soil –moisture and rainfall composite during the break phase of the monsoon with 12 days lead-lag. Day 0 represents the composite of all break days. On day 0, negative soil moisture persists all over India except the northeast India and some Tamilnadu coast. Maximum negative soil moisture anomaly is available over Northwest India and some of central India. There is a similar east west gradient in the soil moisture like the active phase at day 11.The details of soil moisture's role during the monsoon break phase will be discussed in a future study.

To see how an operational model performs during active days, we have done the same analysis with the CFS free long run, shown in figs 12-14. This suggests that the model could not capture the observed features of specific humidity, as shown in fig 8. Similarly, fig 10 shows that the Brunt Vaisala frequency over the north and south boxes of the monsoon core zone, a measure of instability in the atmosphere, has a very different pattern in the model from the observation. So this operational model does not show the pre-conditioning role of soil moisture, and its feedback to rainfall is absent during the active phase of the monsoon. We found that one possible reason could be the improper arrangements of soil type in the model, leading to improper soil moisture–precipitation feedback. We conclude that to get proper feedback between soil moisture and precipitation during the active phase of monsoon in the model, the soil properties of the model could be improved.

**Acknowledgments**

The authors acknowledge the data support from Dr. Suryachandra A. Rao and Dr. Ankur Srivastava from the Indian Institute of Tropical Meteorology, Pune (IITM). Soil moisture data were obtained from the monsoon mission project (https://www.tropmet.res.in/monsoon/monsoon2/). The authors also thank Dr. Ananya Karmakar from India Meteorological Department Pune, Maharashtra (IMD, Pune) for providing the Koppen Classification plot. The authors acknowledge the research and funding support from the Indian Institute of Tropical Meteorology, Pune (IITM), fully supported by the Earth System Science Organization of the Ministry of Earth Sciences (MoES), Govt. of India, and from the Weather and Climate Science for Service Partnership (WCSSP) India, a collaborative initiative between the Met Office, supported by the UK Government's Newton Fund, and the MoES.

**Data availability** The datasets used for this study are available freely from ERA5 reanalysis datasets (Hersbach et al., 2020) (https://www.ecmwf.int/en/forecasts/datasets/reanalysisfor soil –moisture at level 1(0-10cm), level 2 (10-28cm), level 3 (29-78), and level 4 (79-100cm), the same for soil –temperature data at different levels along with evaporation, specific humidity and temperature data for the period 1989-2019. NCEP and (Nayak et al., 2018) datasets are also used for soil–moisture parameters, but the soil–moisture level is different in all the data sets. In Nayak, H. P. et al. 2018 soil –moisture level 1 is 0-7cm. We have denoted this data as MMSM in this study. In addition to the soil moisture data, we used IMD (India Meteorological Department) gridded rainfall data from Pai et al. (2014). The gridded rainfall data are available for the entire Indian subcontinent (6.5N–37.5N and 66.5E–101.5E). We also used political map of India from https://www.mapsofindia.com/ to draw the soil type over Indian region.

**Code availability** The codes used for the study will be available upon request from the corresponding author.

**Declarations**

**Competing interests** The authors declare no competing interests.

# Figure captions

Fig:1 (a) JJAS climatology of soil-moisture at level-1 (SM-1 m*3 /m*3)(shaded) and MSLP (hPa) (contour) from ERA5 data, (b) Active day composite of anomalous SM-1 (shading) with actual MSLP(*100 hPa contour) and dot pattern of anomalous omega(*-10^-2) at 500hPa for the period 1989-2017. Fig 1(c) and 1(d) show the different soil types of India and the Koppen climate classification of the Indian region from 1981-2010.

Fig: 2 (a),(b), and (c) show time-series plot for rainfall, evaporation, and ERA5 SM-1 anomaly during JJAS over Central India for the year 2000. (d) (e) and (f) shows its power spectra pattern during JJAS for the period (1989-2019).

Fig: 3 Correlation of ERA5 soil moisture at different levels with non-filtered IMD rainfall during JJAS (1989-2019) over Central India.

Fig: 4 Lead-lag latitude section of (a) anomalous soil moisture at lev 1 (SM-1) (*0.001 $m^3/m^3$)(shaded) and rainfall(mm/day)(contour) anomalies averaged over 68-96E for the period 1989-2017, on days before (lead; negative) and after (lag; positive) the rainfall maximum at day 0. (b), (c) and (d) the same for different soil levels.

Fig:5 Spatial Distribution of anomalous ERA5 SM-1(*0.01m3/m3) (shaded) in the days before, during and after the peak of the active phase (lag 0). Rainfall(mm/day) is shown as contours.

Fig:6 (a) Spatial pattern of anomalous ERA5 SM-1 (shaded) and rainfall (contours) on the peak active day (day 0), showing the two boxes (north of monsoon core zone (N) and south of monsoon core zone (S). (b) shows the standardized anomaly of SM-1 and rainfall gradient between (N minus S) monsoon core zone in the days before (negative), during, and after (positive) the peak active day 0. (c) and (d) shows the probability distributions of standardized rainfall and soil moisture anomaly over the N and S boxes for days -5 to -1 before the peak active day.

Fig:7 Spatial Distribution of Evaporation (shaded) and MSLP (contour) in the days leading up to the peak active phase (day 0).

Fig:8 Vertical profile of Specific humidity anomaly ($10^{-4}$ kg/kg) gradient over (N minus S) monsoon core zone during active period (peak at day 0). Positive color indices (1,2,3,4,5) represent days after day 0 and negative color indices (-1,-2,-3,-4,-5) represent days before day 0.

Fig:9 Spatial Distribution of anomalous MSE (x100 J/kg) averaged over 1000-150hPa (shaded) in the 12 days leading up to and after the peak active phase (peak at lag 0). Rainfall (mm/day) is shown as contours.

Fig:10 Brunt Vaisala frequency over North and south monsoon core zone during days before and after the peak active day (day 0) from ERA5 datasets and CFS free run.

Fig:11 : Proposed mechanism of soil moisture response over monsoon core zone.

Fig:12 Spatial Distribution of anomalous SM-1(m3/m3) (shaded) from CFS free run during active phase (peak at lag 0). Rainfall(mm/day) is shown as contours.

Fig:13 Standardized SM-1 and rainfall gradient over (N minus S) monsoon core zone during days before and after the peak active day (day 0) from CFS free run.

Fig:14 Vertical profile of Specific humidity anomaly ($10^{-4}$ kg/kg ) over (N minus S) monsoon core zone from CFS free run during active period (peak at day 0).Positive color indices (1,2,3,4,5) represent days after day 0 and negative color indices (-1,-2,-3,-4,-5) represent days before day 0.

# Supplementary figures

Fig-S1(a) Time-series plot for soil-moisture (SM-1) anomaly in MMSM (Nayak et al., 2018) during JJAS over Central India for the year 2000. (b) Power spectra pattern for MMSM during JJAS for the period (1989-2017).

Fig: S2 Correlation of ERA5 soil-moisture at different levels (level-1,level-2,level-3and level-4 respectively) with filtered (20-100 days) IMD rainfall during JJAS (1989-2019) over Central India.

Fig: S3 Spatial Distribution of anomalous SM-1(m3/m3) (shaded) from MMSM (Nayak et al., 2018) in the days before, during and after the peak of the the active phase (peak at lag 0). IMD Rainfall (mm/day) is shown as contours.

Fig :S4 Actual value of rainfall composite during active days(1989-2019) .Black box shows the north box we have selected for this study.

Fig :S5 Actual value of rainfall composite during active days over North box selected in the above figure.

Fig:S6 Spatial Distribution of anomalous SM-1(*0.01m3/m3) (shaded) during break phase (lag 0).Rainfall(mm/day) is shown as contours .

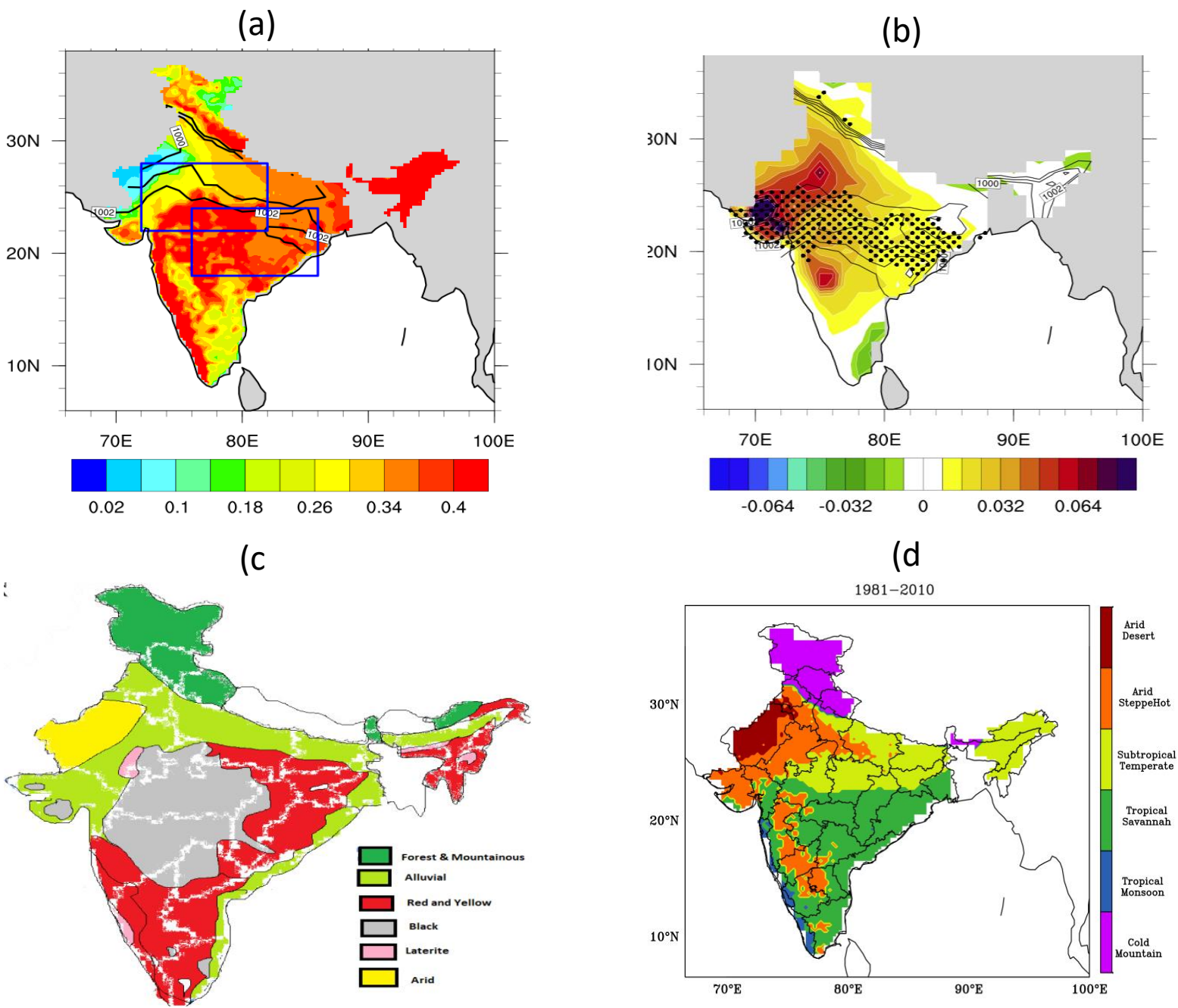

Fig:1 (a) JJAS climatology of soil-moisture at level-1 (SM-1 m*3 /m*3)(shaded) and MSLP (hPa) (contour) from ERA5 data,(b) Active day composite of anomalous SM-1 (shading) with actual MSLP(*100 hPa contour) and dot pattern of anomalous omega(*-10^-2) at 500hPa for the period 1989-2017. Fig 1(c) and 1(d) show the different soil types of India and the Koppen climate classification of the Indian region from 1981-2010.

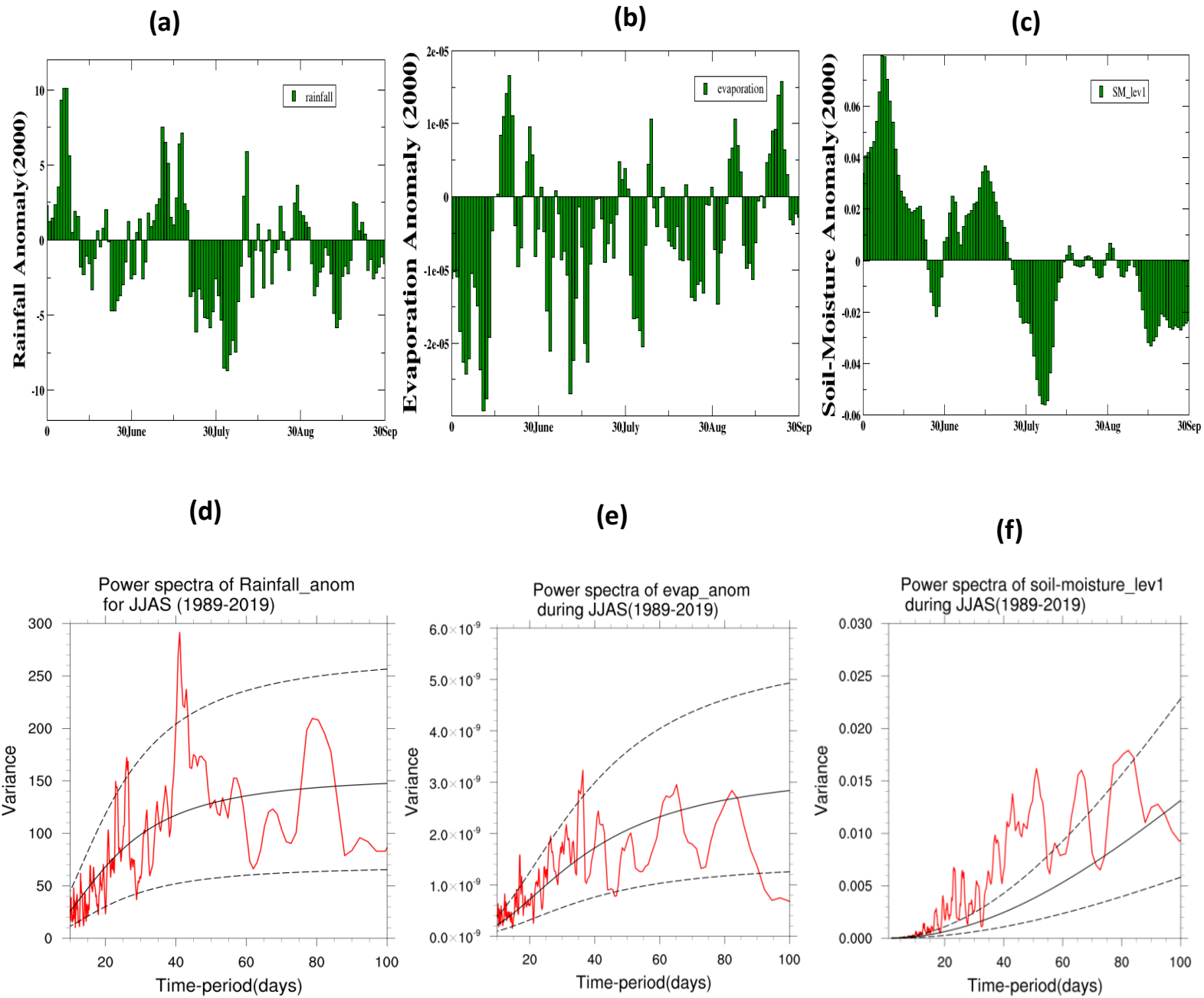

**Fig-2(a) ,(b) and (c ) represent the time-series plot for rainfall, evaporation and SM-1 anomaly during JJAS over Central India for the year 2000.(d) ,(e) and (f) shows its power spectra pattern during JJAS for the period (1989-2019) .**

(a)

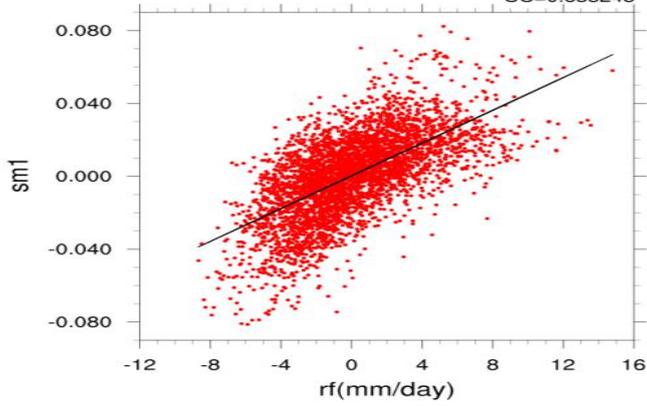

(b)

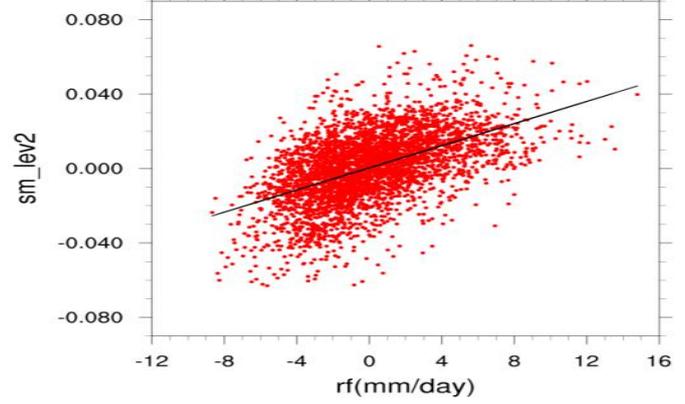

(c)

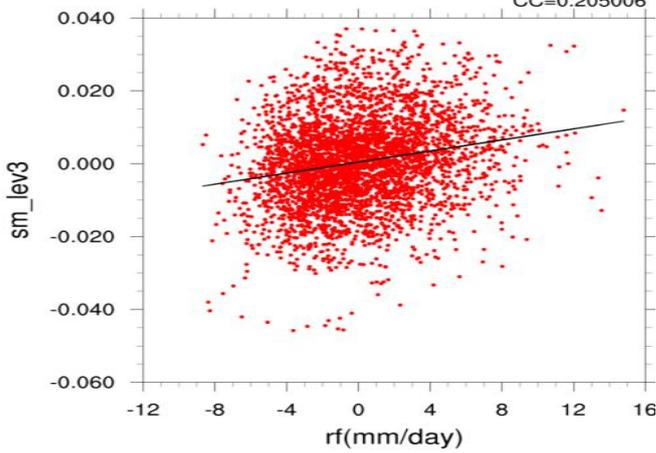

(d)

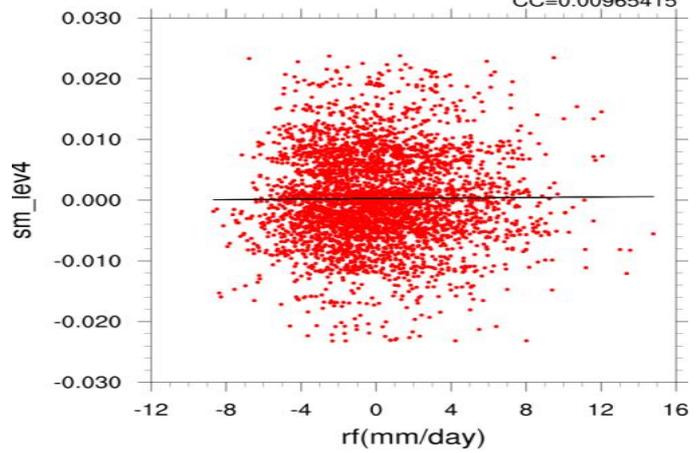

**Fig:3 Correlation with soil-moisture at different levels (level-1,level-2,level-3 and level-4 respectively) with non-filtered rainfall during JJAS(1989-2019) over Central India.**

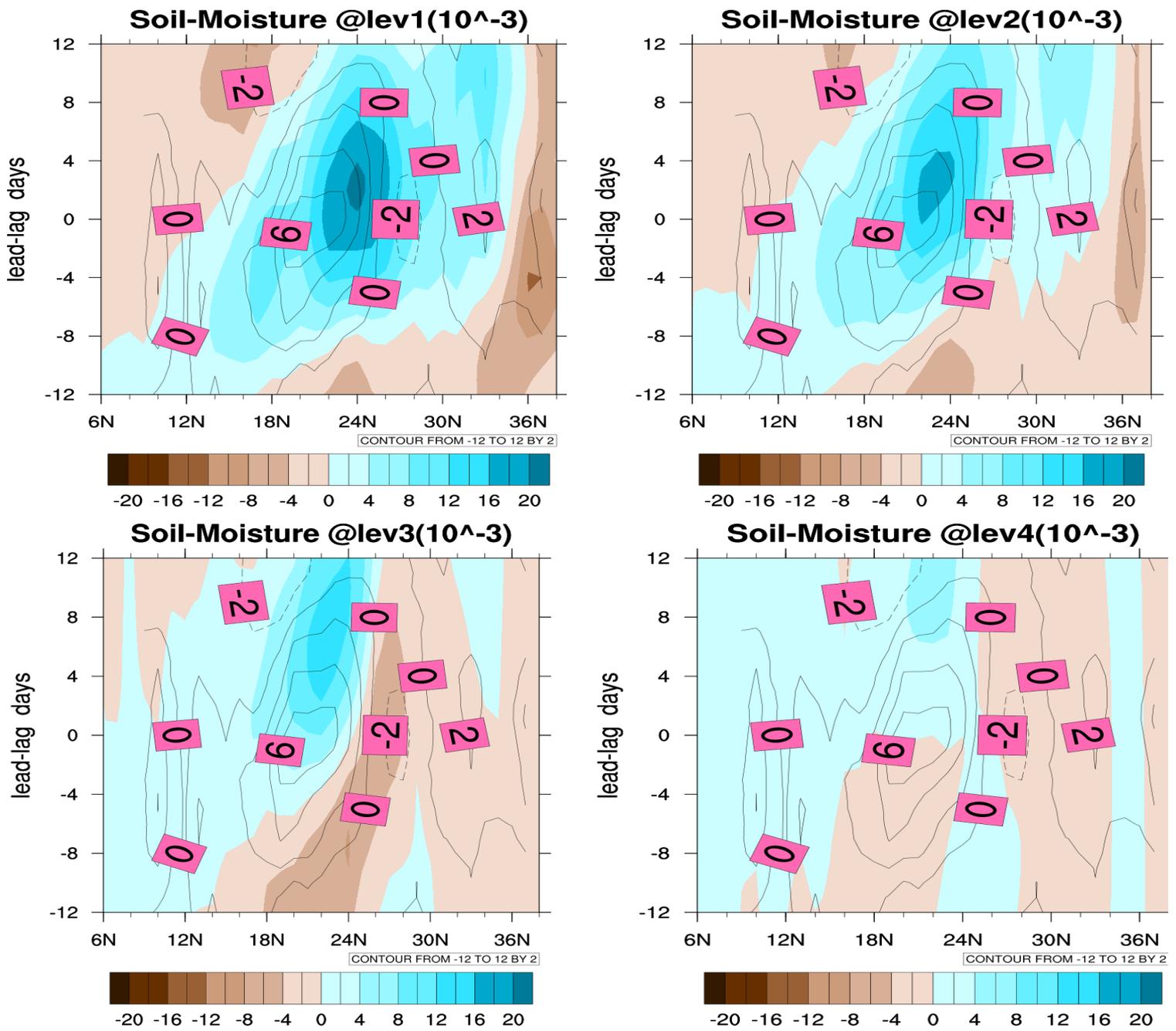

**Fig:4** Lead-lag latitude section of (a) anomalous soil moisture at lev 1 (SM-1) (*0.001 m$^3$/m$^3$)(shaded) and rainfall(mm/day)(contour) anomalies averaged over 68-96E for the period 1989-2017, on days before (lead; negative) and after (lag; positive) the rainfall maximum at day 0. (b), (c) and (d) the same for different soil levels.

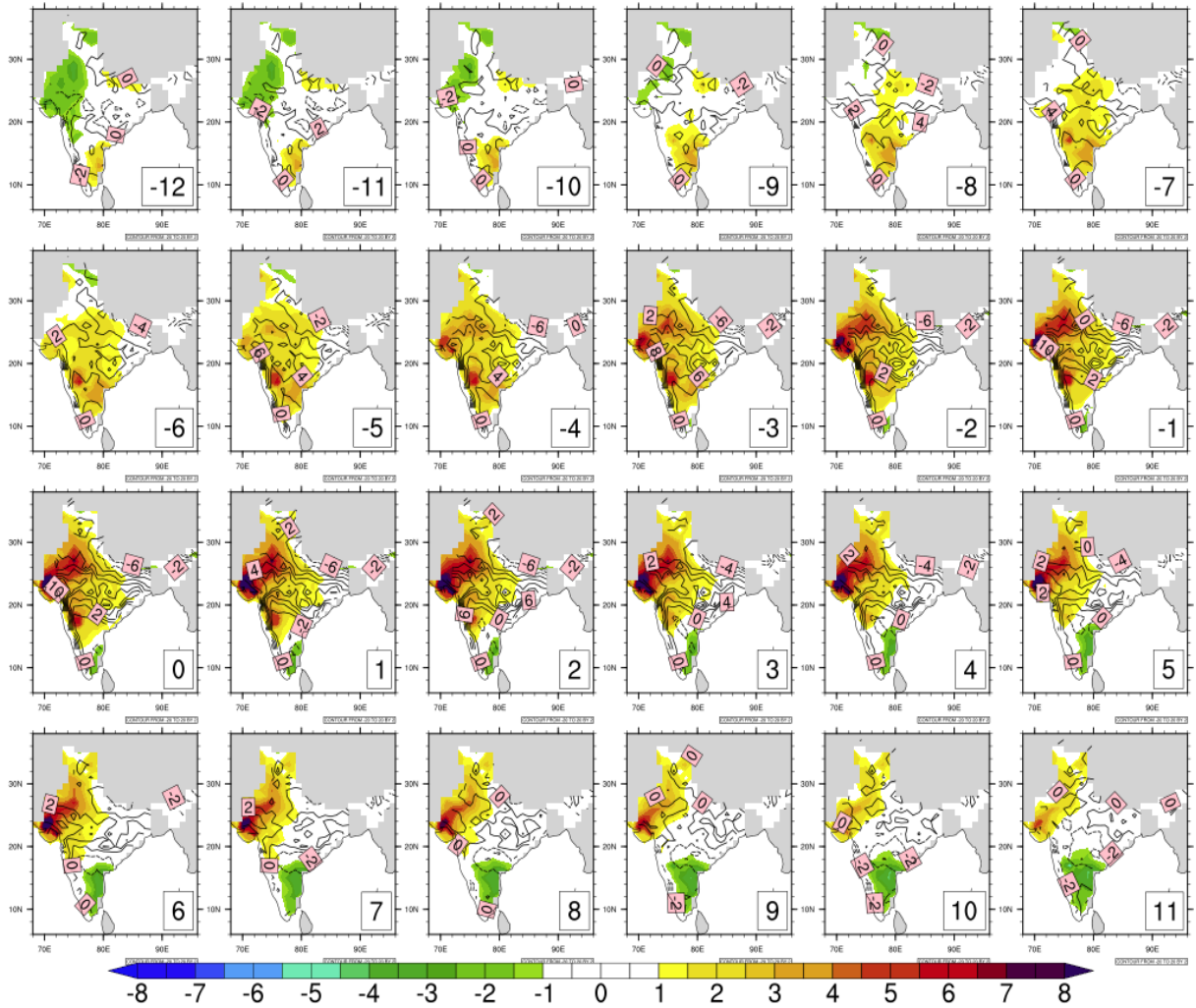

**Fig:5 Spatial Distribution of anomalous SM-1(*0.01 m3/m3) (shaded) during active phase (lag 0).Rainfall(mm/day) is shown as contours .**

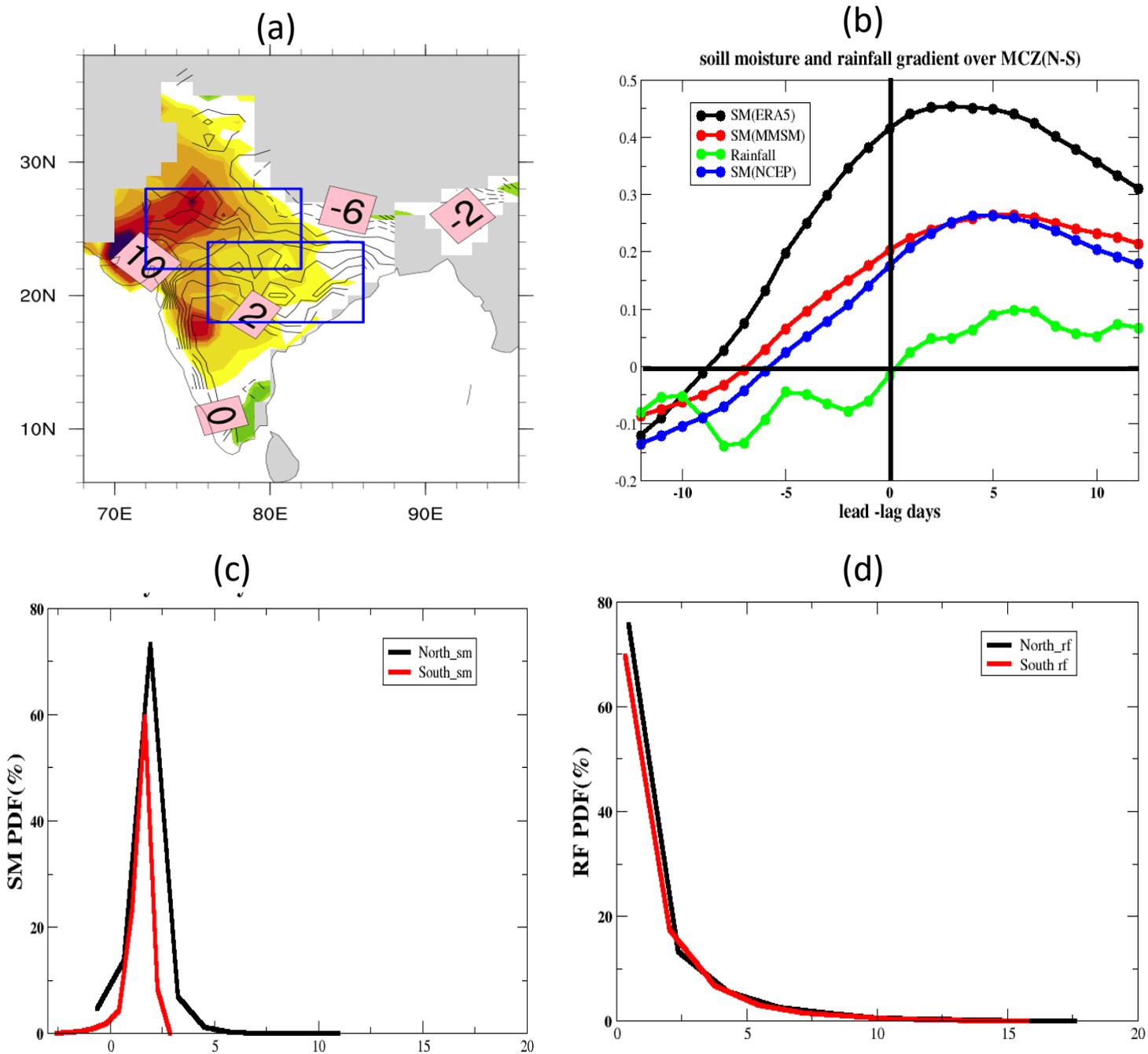

Fig: 6 (a) Represents the spatial shift of anomalous SM-1 over the two boxes (north of monsoon core zone (N) and south of monsoon core zone(S). (b) plotted the standardized anomaly of SM-1 and rainfall gradient over (N-S) monsoon core zone during active days.(c)and (d) shows soil-moisture and rainfall pdf over north and south MCZ.

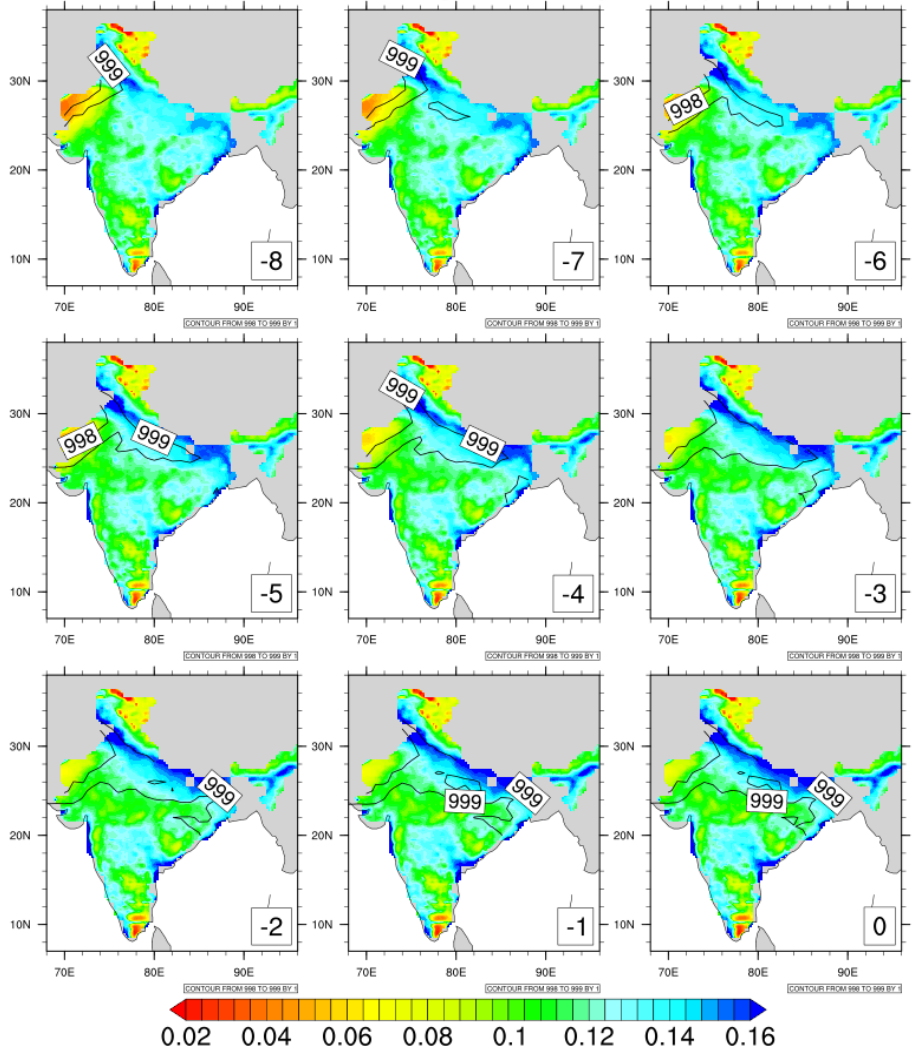

**Fig: 7 Spatial Distribution of Evaporation (shaded) and MSLP (contour) during active phase.**

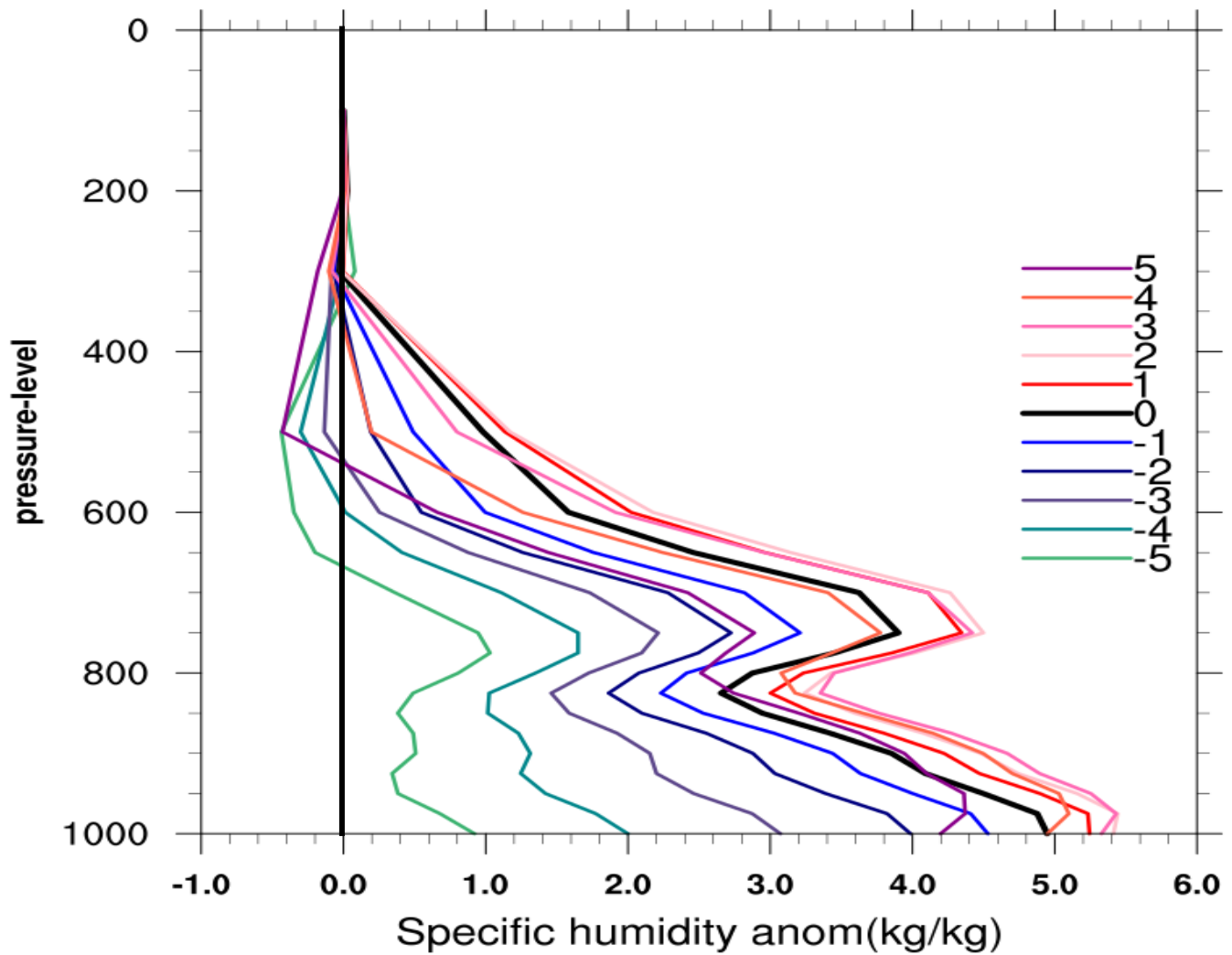

**Fig:8** Vertical profile of Specific –humidity anomaly (10^-4 kg/kg ) over (N-S) monsoon core zone during active days(day 0) .Where positive color index (1,2,3,4,5) represent for lead from days 0 and (-1,-2,-3,-4,-5) for lag days from day0.

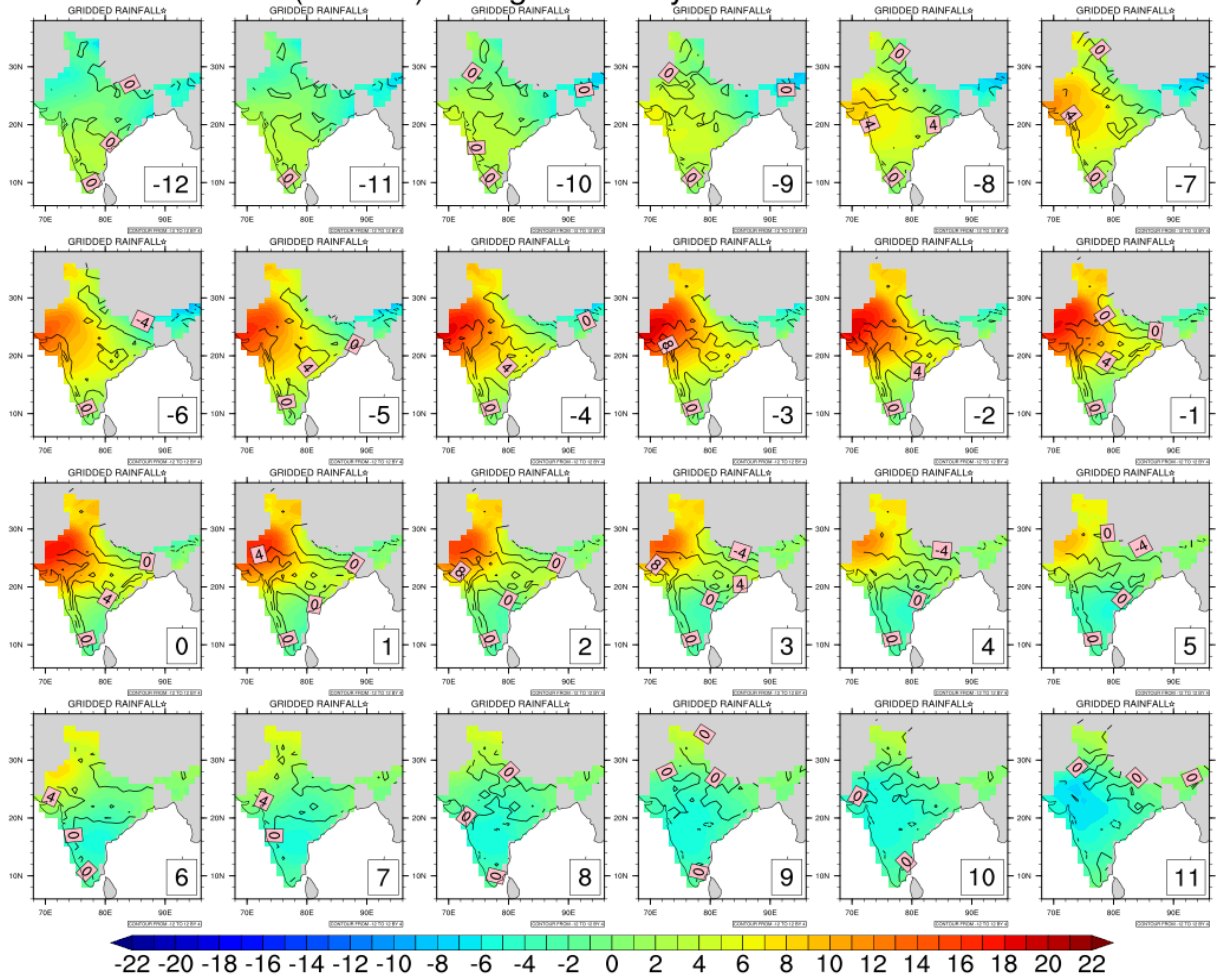

Fig:9 Spatial Distribution of anomalous MSE(x100 J/kg) (shaded) during active phase (lag 0).Rainfall(mm/day) is shown as contours .

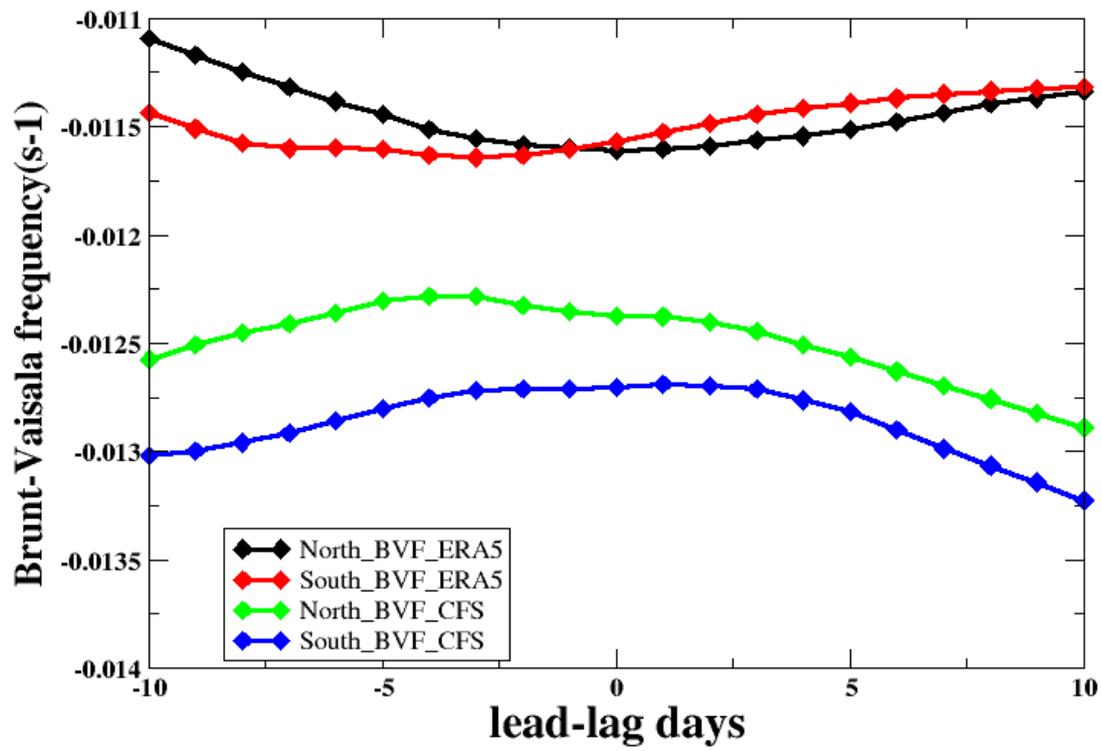

Fig 10 Brunt Vaisala frequency over North and south monsoon core zone during active days(day 0) from ERA5 datasets and CFS free run.

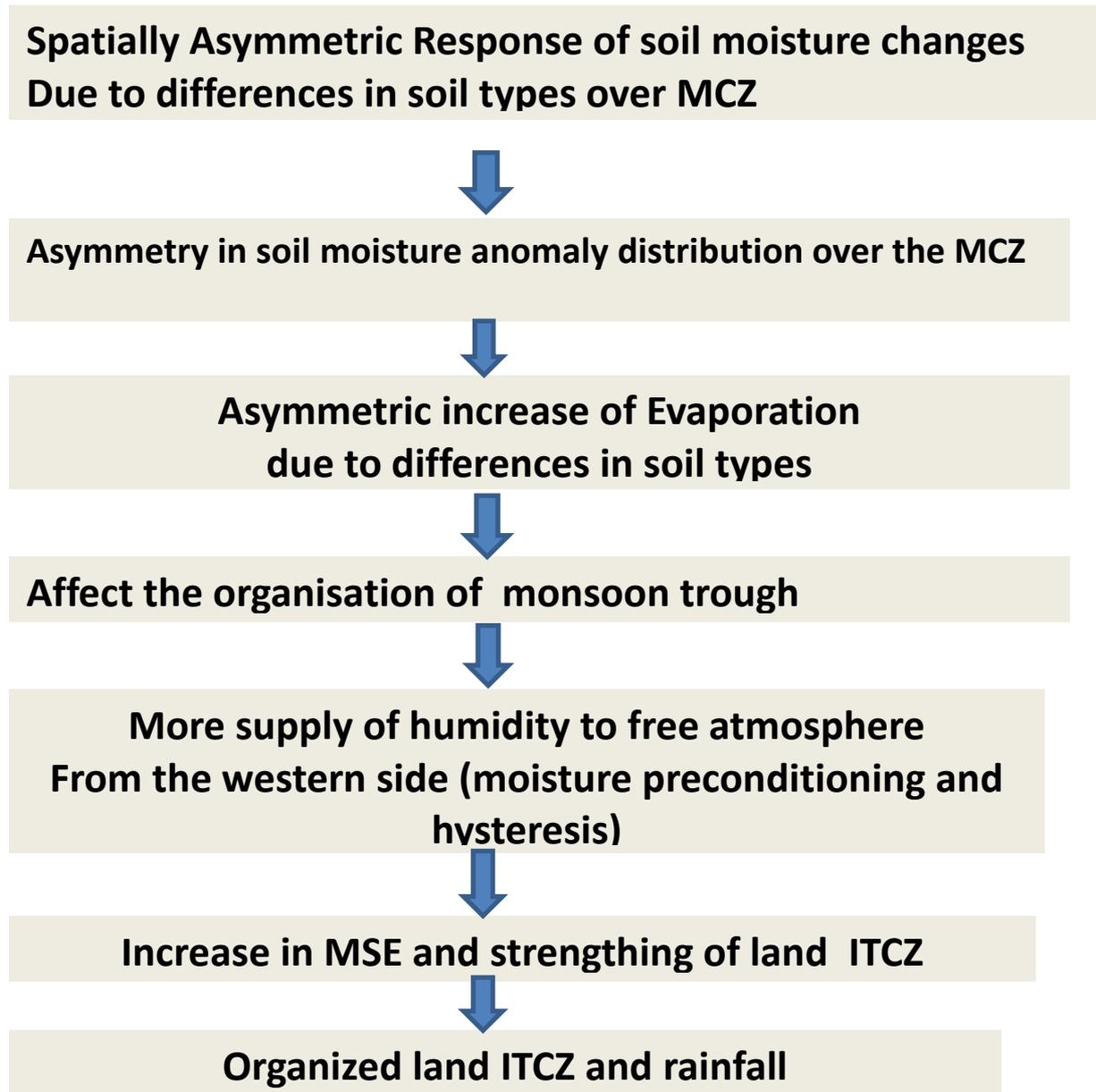

fig.11: Proposed mechanism of soil moisture response over MCZ.

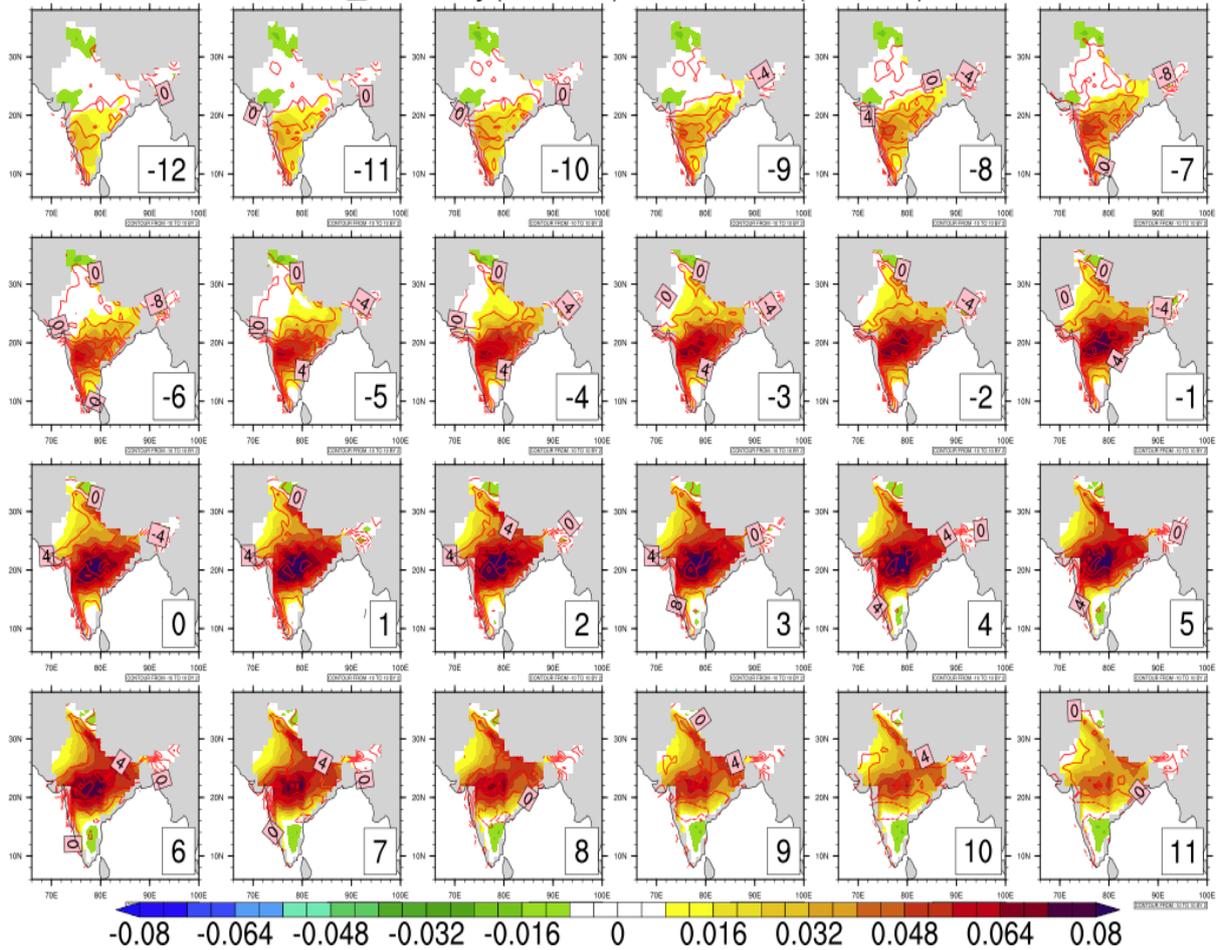

**Fig:12 Spatial Distribution of anomalous SM-1(m3/m3) (shaded) during active phase (lag 0).Rainfall(mm/day) is shown as contours from CFS free run .**

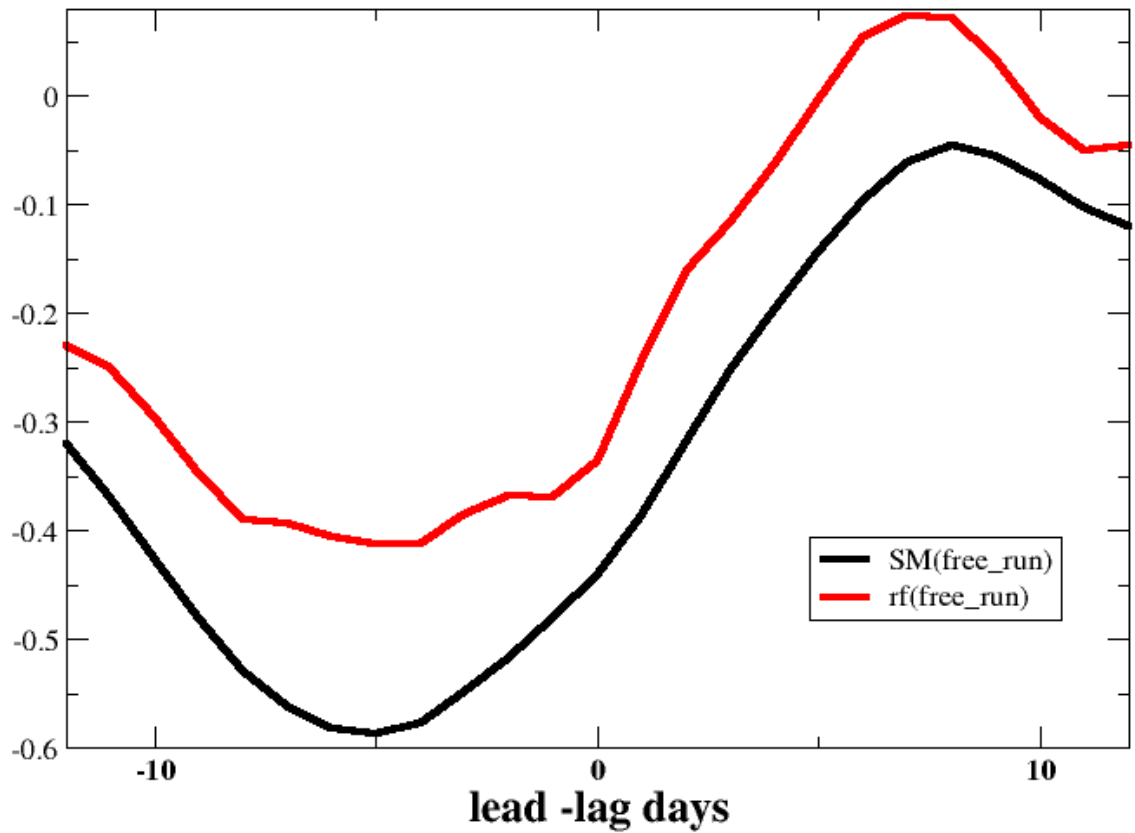

**Fig:13 Plotted the standardize SM-1 and rainfall gradient over (N-S) monsoon core zone during active days from CFS free run .**

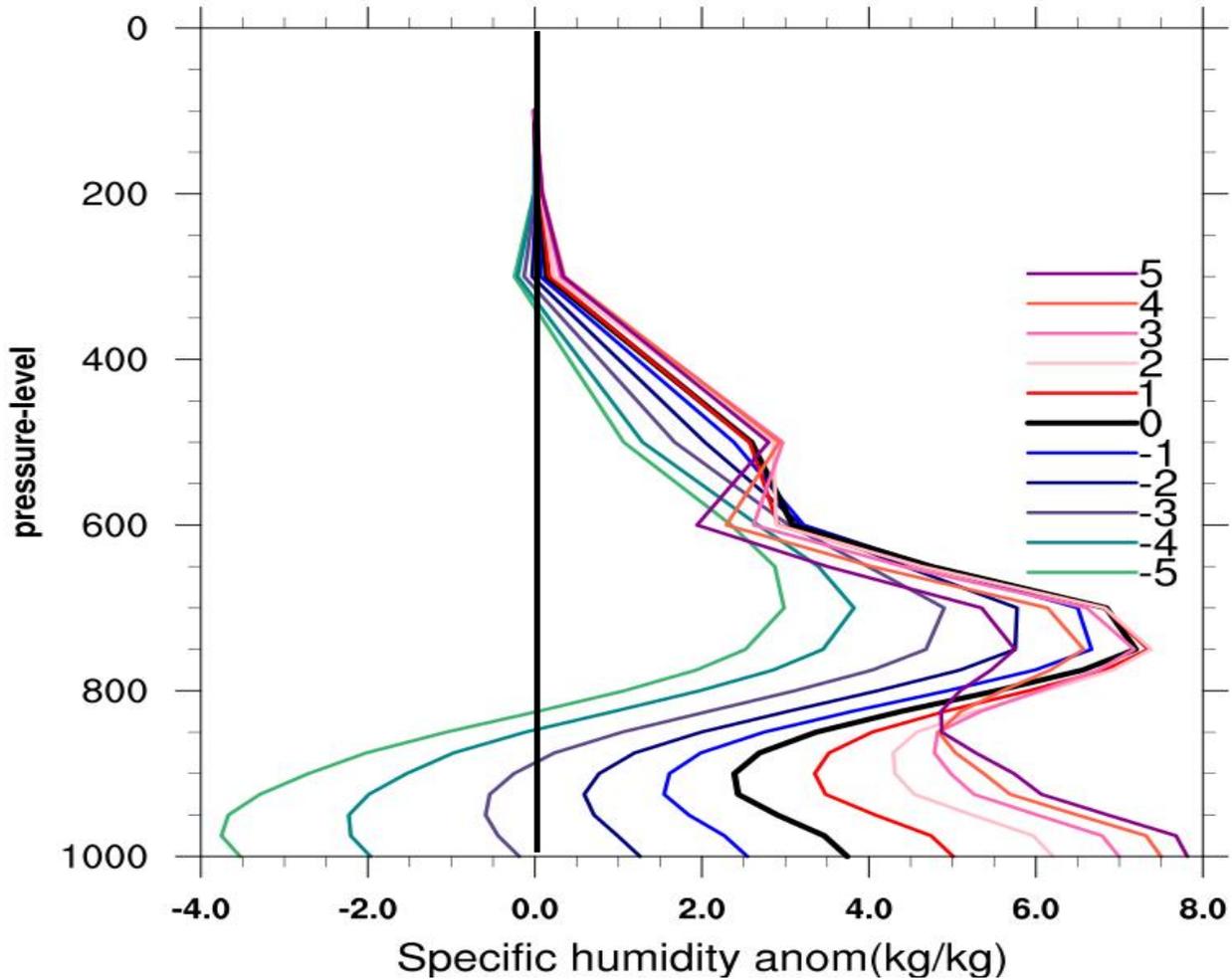

**Fig:14** Vertical profile of Specific –humidity anomaly (10^-4 kg/kg ) over (N-S) monsoon core zone during active days(day 0) .Where positive color index (1,2,3,4,5) represent for lead from days 0 and (-1,-2,-3,-4,-5) for lag days from day0 form CFS free run.

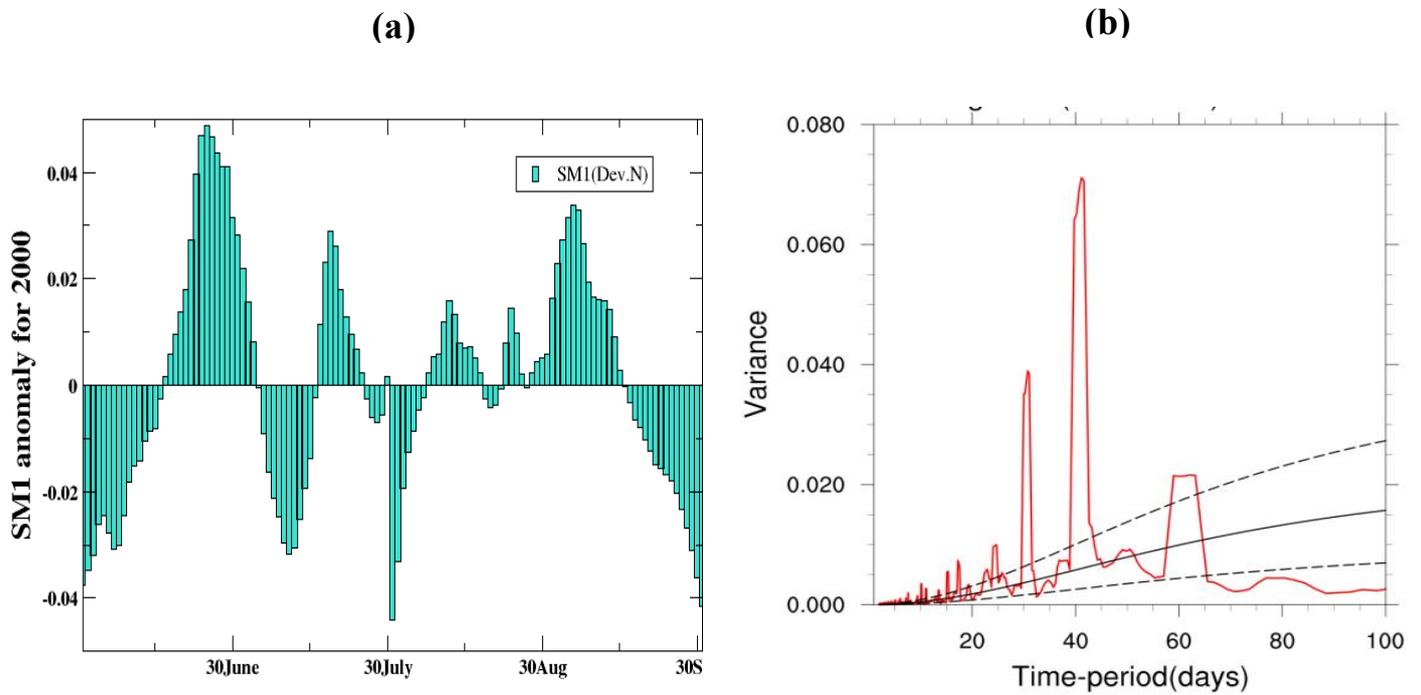

**Fig-S1(a) represents the time-series plot for soil-moisture (SM-1) anomaly during JJAS over Central India for the year 2000. (b) Shows its power spectra pattern during JJAS for the period (1989-2017) from Nayak et al., 2018 datasets.**

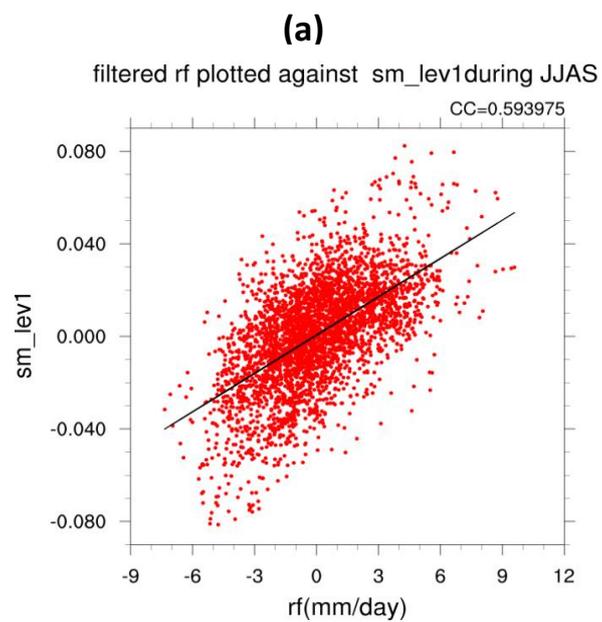 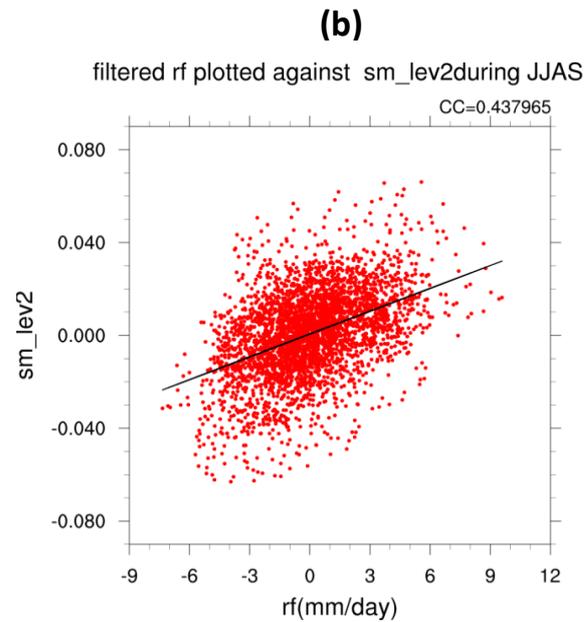
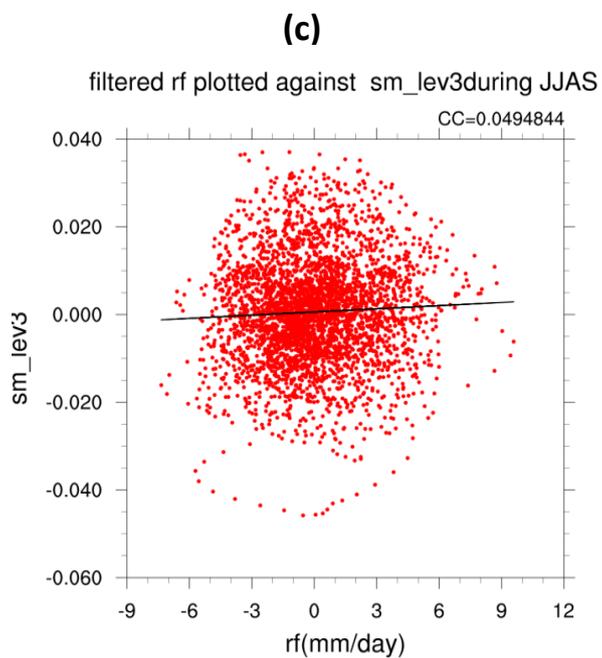 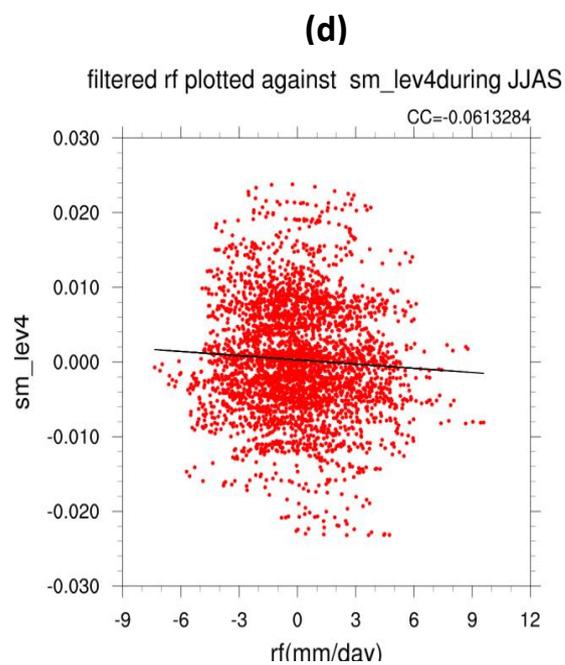

Fig:S 2 Correlation with soil-moisture at different levels (level-1,level-2,level-3and level-4 respectively) with filtered rainfall during JJAS(1989-2019) over Central India.

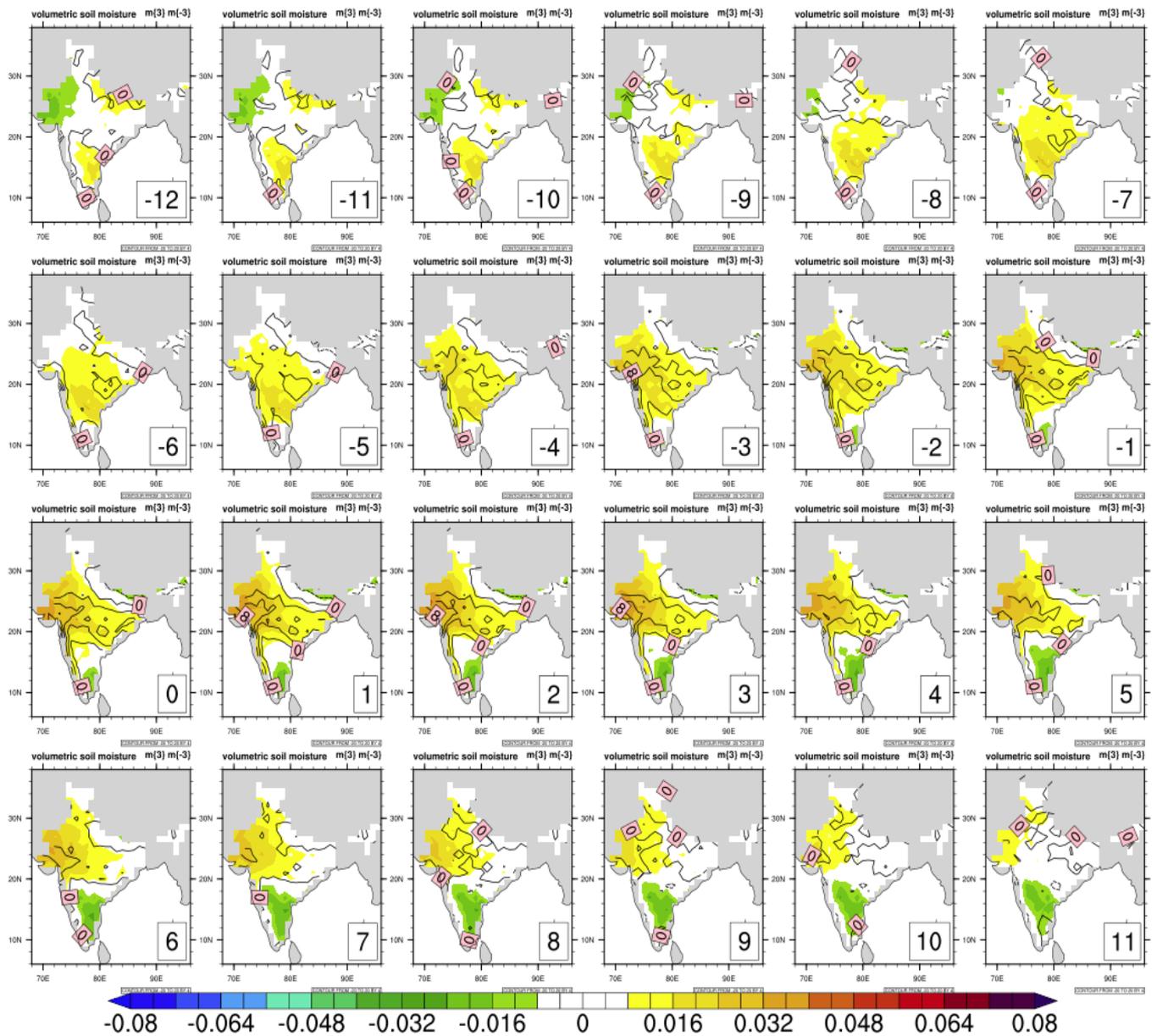

Fig:S3 Spatial Distribution of anomalous SM-1(m3/m3) (shaded) during active phase (lag 0).Rainfall(mm/day) is shown as contours from Nayak et al., 2018 datasets.

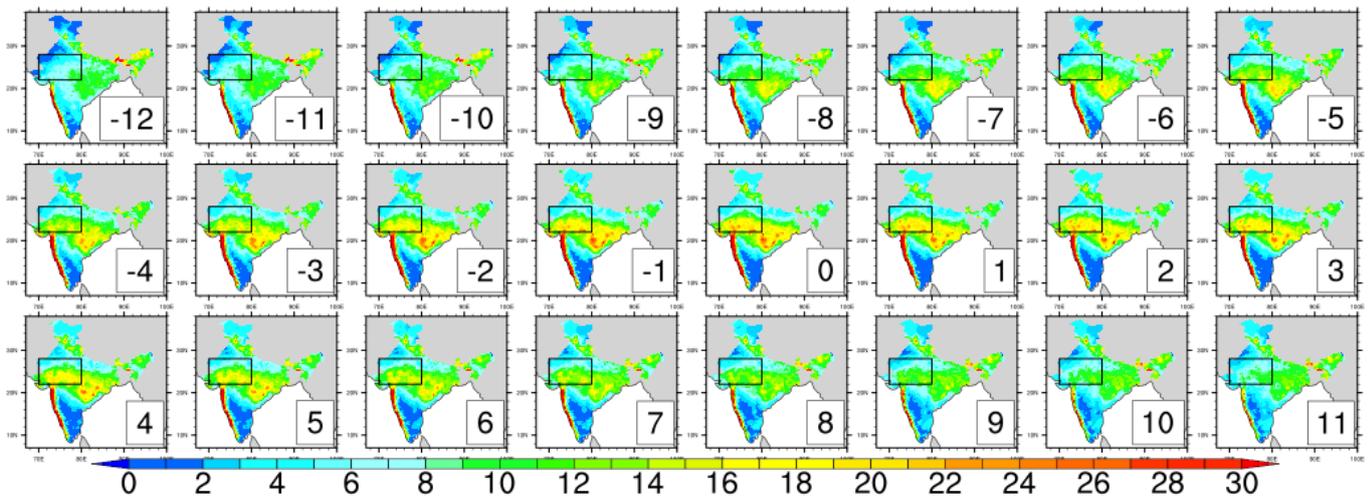

**Fig :S4 Actual value of rainfall composite during active days(1989-2019) .Black box shows the north box we have selected for this study.**

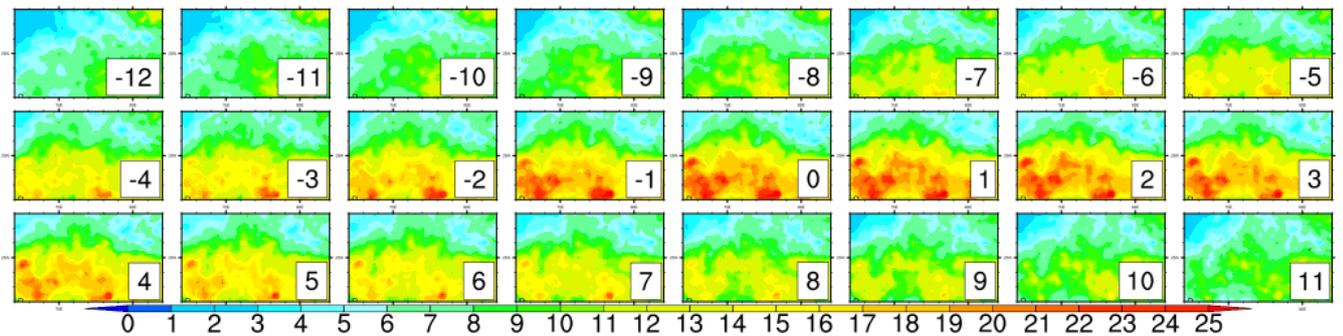

**Fig :S5 Actual value of rainfall composite during active days over North box selected in the above figure.**

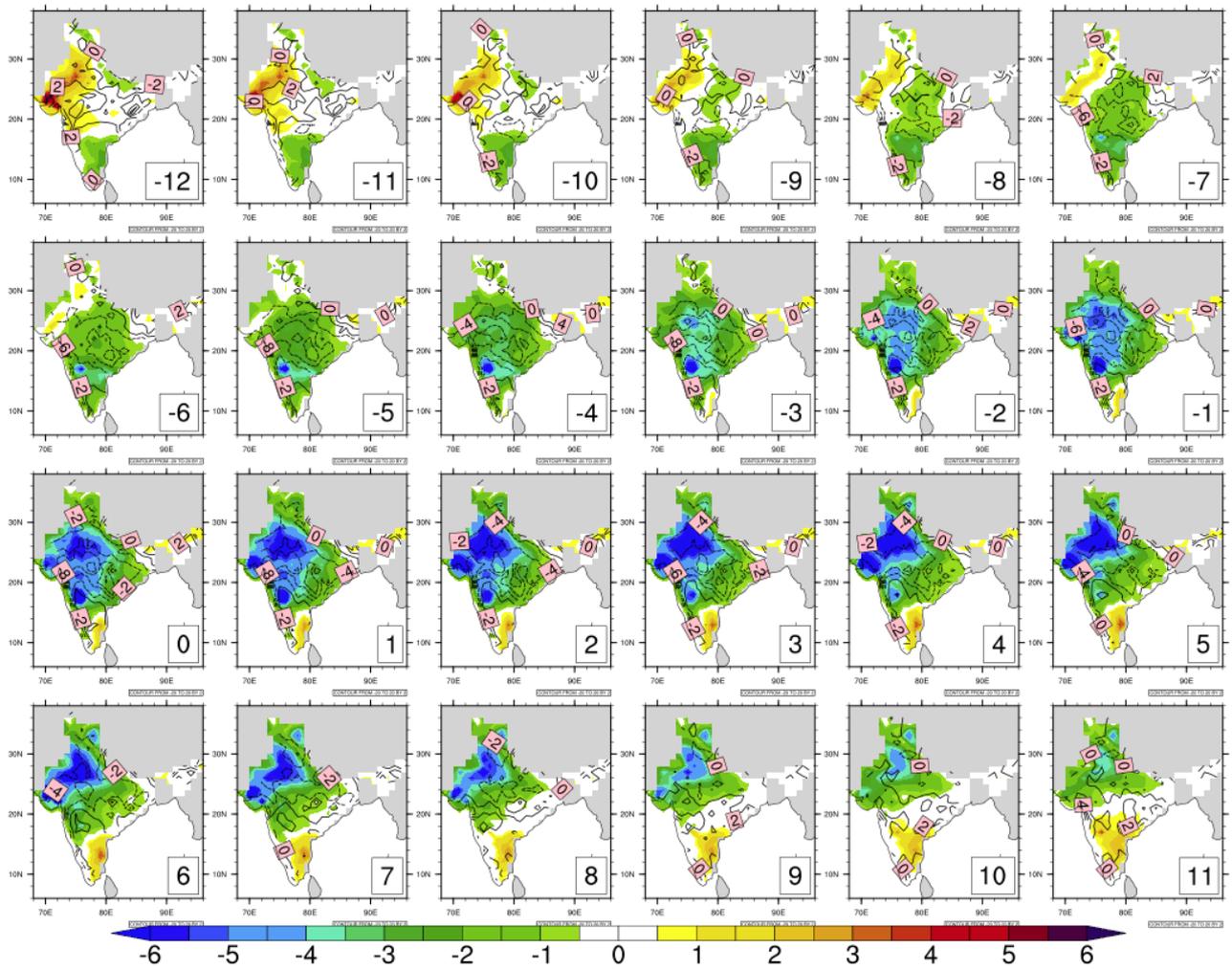

**Fig:S6** Spatial Distribution of anomalous SM-1(m3/m3) (shaded) during break phase (lag 0).Rainfall(mm/day) is shown as contours .